\documentclass[sigconf]{acmart}

\usepackage{booktabs} 
\usepackage{multirow}
\usepackage{enumitem}
\usepackage{balance}
\usepackage[ruled]{algorithm2e}
\usepackage[flushleft]{threeparttable}
\usepackage{bm}
\usepackage{xcolor}
\usepackage{color}

\newcommand{\eat}[1]{}

\copyrightyear{2022} 
\acmYear{2022} 
\setcopyright{acmcopyright}\acmConference[SIGIR '22]{Proceedings of the 45th International ACM SIGIR Conference on Research and Development in Information Retrieval}{July 11--15, 2022}{Madrid, Spain}
\acmBooktitle{Proceedings of the 45th International ACM SIGIR Conference on Research and Development in Information Retrieval (SIGIR '22), July 11--15, 2022, Madrid, Spain}
\acmPrice{15.00}
\acmDOI{10.1145/3477495.3532026}
\acmISBN{978-1-4503-8732-3/22/07}

\acmArticle{4}

\begin{document}
\fancyhead{}

\title{Multi-Level Interaction Reranking with User Behavior History}

\author{Yunjia Xi}
\authornote{Both authors contributed equally to this research.}
\email{xiyunjia@sjtu.edu.cn}
\affiliation{%
  \institution{Shanghai Jiao Tong University}
  \city{Shanghai}
  \country{China}
}

\author{Weiwen Liu}
\authornotemark[1]
\email{liuweiwen8@huawei.com}
\affiliation{%
  \institution{Huawei Noah's Ark Lab}
  \city{Shenzhen}
  \country{China}
}

\author{Jieming Zhu}
\email{jamie.zhu@huawei.com}
\affiliation{%
  \institution{Huawei Noah's Ark Lab}
  \city{Shenzhen}
  \country{China}
}

\author{Xilong Zhao}
\email{zhaoxilong@sjtu.edu.cn}
\affiliation{%
  \institution{Shanghai Jiao Tong University}
  \city{Shanghai}
  \country{China}
}

\author{Xinyi Dai}
\email{xydai@apex.sjtu.edu.cn}
\affiliation{%
  \institution{Shanghai Jiao Tong University}
  \city{Shanghai}
  \country{China}
}

\author{Ruiming Tang}
\email{tangruiming@huawei.com}
\affiliation{%
  \institution{Huawei Noah's Ark Lab}
  \city{Shenzhen}
  \country{China}
}

\author{Weinan Zhang}
\authornote{Corresponding author.}
\email{wnzhang@sjtu.edu.cn}
\affiliation{%
  \institution{Shanghai Jiao Tong University}
  \city{Shanghai}
  \country{China}
}

\author{Rui Zhang}
\email{rayteam@yeah.net}
\affiliation{%
  \institution{ruizhang.info}
}

\author{Yong Yu}
\authornotemark[2]
\email{yyu@sjtu.edu.cn}
\affiliation{%
  \institution{Shanghai Jiao Tong University}
  \city{Shanghai}
  \country{China}
}


\begin{abstract}
As the final stage of the multi-stage recommender system (MRS), reranking directly affects users' experience and satisfaction, thus playing a critical role in MRS. Despite the improvement achieved in the existing work, three issues are yet to be solved. First, users' historical behaviors contain rich preference information, such as users' long and short-term interests, but are not fully exploited in reranking. Previous work typically treats items in history equally important, neglecting the dynamic interaction between the history and candidate items. Second, existing reranking models focus on learning interactions at the item level while ignoring the fine-grained feature-level interactions. Lastly, estimating the reranking score on the ordered initial list before reranking may lead to the early scoring problem, thereby yielding suboptimal reranking performance. To address the above issues, we propose a framework named Multi-level Interaction Reranking (MIR). MIR combines low-level \textbf{cross-item interaction} and high-level \textbf{set-to-list interaction}, where we view the candidate items to be reranked as a \textit{set} and the users' behavior history in chronological order as a \textit{list}. We design a novel SLAttention structure for modeling the set-to-list interactions with personalized long-short term interests. Moreover, \textbf{feature-level interactions} are incorporated to capture the fine-grained influence among items. 
We design MIR in such a way that any permutation of the input items would not change the output ranking, and we theoretically prove it. Extensive experiments on three public and proprietary datasets show that MIR significantly outperforms the state-of-the-art models using various ranking and utility metrics.

\end{abstract}


\maketitle

\section{Introduction}
Multi-stage recommender systems (MRS) are widely adopted by many of today's large online platforms, such as Google \cite{seq2slate}, YouTube \cite{wilhelm2018practical}, LinkedIn \cite{geyik2019fairness}, and Taobao \cite{prm}. These systems generate recommendations in multiple stages, including matching, ranking, and reranking. Each stage narrows down the relevant items with a computationally more expensive but more accurate model compared to the previous stage \cite{Jiri2021On}. Reranking, as the final stage, further scales down the candidate items, refines the ranking lists from the previous ranking stage by considering listwise context, and serves the recommendation results to users. The goal of reranking is to optimize the total utility (e.g., number of clicks or overall revenue) of the reranked lists. The quality of reranking directly affects users’ experience and satisfaction, and thus plays a critical role in MRS. Despite some success achieved in existing reranking research \cite{dlcm, prm, setrank, GSF, miDNN}, 
there still exist the following major limitations.

\begin{figure}
    \centering
    \includegraphics[width=\columnwidth]{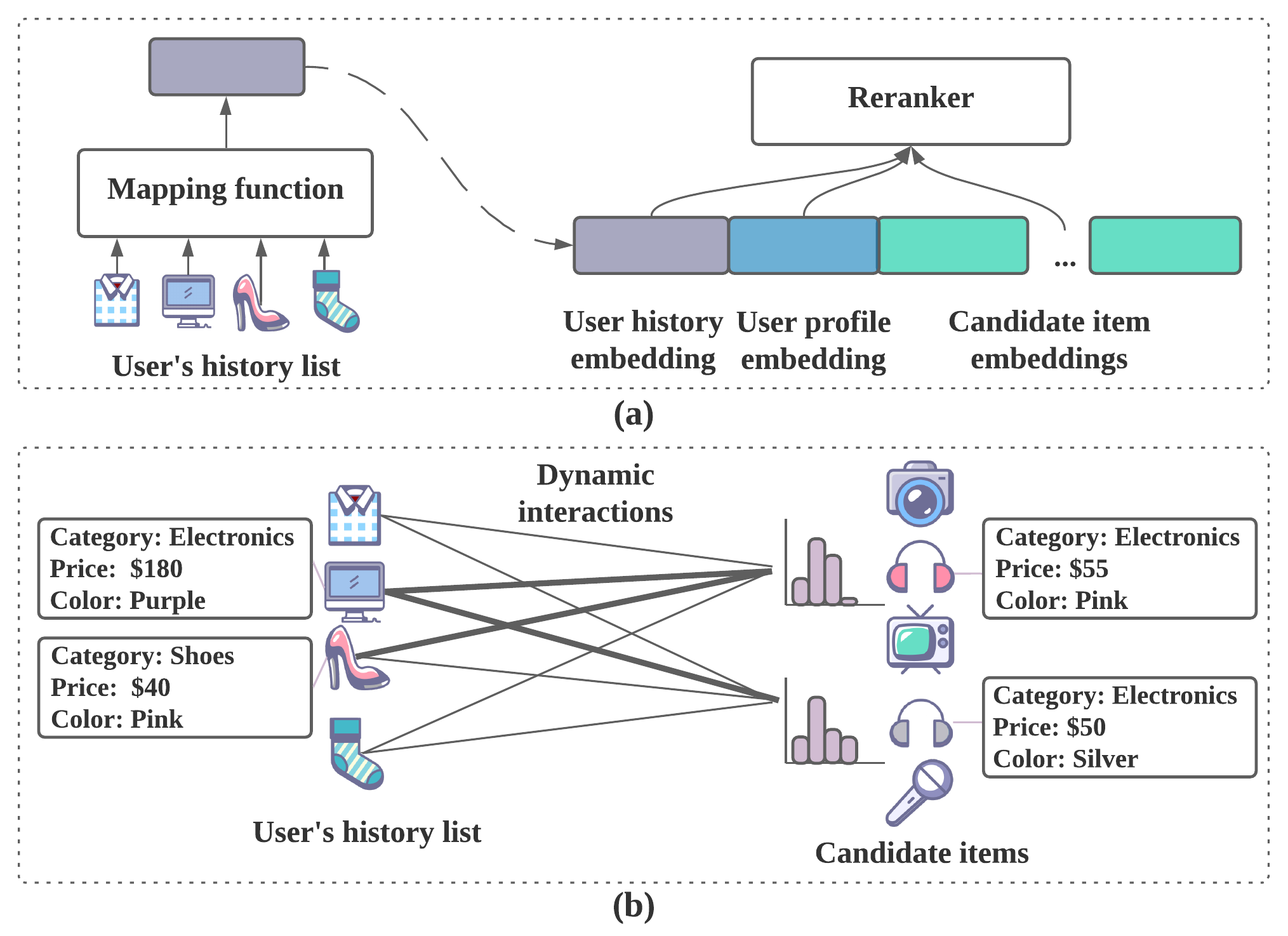}
    \caption{The usage of users' history lists. (a) The users' history list is independent of the candidate items in the reranking list. (b) The users' history list dynamically interacts with the candidate items.}
    \label{fig:compare_intro}
\end{figure}

Firstly, in reranking, users' behavior history (e.g., recent clicks, purchases, and downloads) contains rich preference information and helps understand users' personalized tastes. However, existing reranking models \cite{prm,feng2021revisit,feng2021grn} generally map items from users' history lists into a low dimensional embedding vector as extra user features, independent of what the candidates are, as shown in Figure~\ref{fig:compare_intro}(a). This approach is insufficient to capture personalized preferences due to the following reasons: (i) The items in the users' history lists contribute differently when reranking different candidate items. For example, in Figure~\ref{fig:compare_intro}(b), a user clicks a recommended headphone in the reranking list mostly because she bought a laptop last week, rather than what kind of shoes she purchased in the past. Simply mapping the historical items to a history embedding vector and treating each history item equally important does not reflect users' dynamic preference with regard to different candidate items. (ii) Users’ interests are intrinsically dynamic and evolving over time. In particular, users have both relatively stable long-term interests that are repetitive in a long time range, and short-term interests that are frequently changing. Existing reranking models overlook such long- and short-term chronological patterns.

Secondly, previous reranking models focus on learning interactions between items  \cite{prm, dlcm, setrank, miDNN, GSF} while ignoring the feature-level interactions within items. Nevertheless, the interactions between the features of items are also significant. As the example illustrated in Figure \ref{fig:compare_intro}(b), the laptop in the history list tends to have a major effect on the candidate items that have the same \texttt{Category} feature, i.e., \texttt{electronics}. In addition, when considering the \texttt{Color} feature, the \texttt{pink} shoes in the history list may also have a positive influence on the \texttt{pink} headphones in the candidate set, since the user may favor the pink color more. This motivates us to model the feature interactions between history items and candidate items so as to capture the influences between items at a fine-grained level.

Thirdly, many existing methods~\cite{dlcm, prm, seq2slate} estimate reranking scores based only on the context of the input initial ranking. However, the reranking operation (sorting the items according to the estimated reranking scores) changes the context of the initial list to the reranked list, leading to a different distribution of utility \cite{feng2021revisit, xi2021contextaware, feng2021grn, setrank}. The reranked list obtained by this strategy is sub-optimal and does not necessarily bring the maximum utility, which is what we call \textit{the early scoring problem}. For example, given an initial ranking (A, B, C), a reranking model outputs the corresponding scores of (0.46, 0.55, 0.3) based on the context of the initial ranking, and then sorts the items to get (B, A, C). However, 0.46 represents the score when A is placed at position 1, before B and C. Once we place A at position 2, its context will change, making the previous estimation imprecise and the item
arrangements sub-optimal. Currently, there are two solutions to this problem: a generator-evaluator approach and a permutation-invariant modeling approach. The former \cite{feng2021grn, feng2021revisit, xi2021contextaware} employs a generator to output possible rankings of candidate items and an evaluator to select the optimal ranking result. The latter \cite{setrank} is a more computationally efficient approach, which takes candidate items as an unordered set and aims to design a permutation-invariant model to find the best ranking irrespective of the permutation of input.

To address the above issues, we propose a model named Multi-level Interaction Reranking (MIR). We view the candidate items to be reranked as an unordered \textit{item set} and users' historical behaviors as a \textit{list} in chronological order. MIR consists of lower-level \textbf{cross-item interactions} within the candidate set and within the history list respectively, as well as higher-level \textbf{set-to-list interactions} (set2list) between them. We present a unique \textit{SLAttention} structure specially designed for capturing the set2list interactions. Two affinity matrices are involved in SLAttention to simultaneously extract \textbf{item- and feature-level interactions} between candidate items and historical items, respectively. Moreover, a personalized time decay factor is further employed to learn long-short term sequential dependencies. MIR is also designed to be permutation-invariant, which is insensitive to the permutations of the input. The main contributions of this paper are summarized as follows: 
\begin{itemize}
    \item We identify the importance of modeling users' dynamic and personalized interests from users' behavior history for reranking. We propose the MIR model, consisting of low-level cross-item interaction within candidate set or history list, and high-level set2list interaction between them. We also introduce feature-level interactions to model fine-grained influences between items. To the best of our knowledge, this is the first work to learn explicit feature-level interactions in reranking.
    \item We propose a novel SLAttention structure for set2list interaction, which models the long-term and short-term interests from users' behavior history and achieves a permutation-invariant reranking to the candidate set. We also theoretically analyze the permutation-invariant property.
    \item Extensive experiments on three public and proprietary datasets show that our MIR outperforms the state-of-the-art baselines in terms of ranking metrics and utility metrics.
\end{itemize}


\section{Related Work}
This section reviews studies on learning to rank and reranking.
\subsection{Learning to Rank}
In MRS, the reranking stage is built on the initial rankings given by the prior ranking stage. Learning to rank that applies machine learning algorithms is one of the most widely used methods in the ranking stage.
According to the loss function, it can be broadly classified into pointwise \cite{mcrank,GBM,prank}, pairwise \cite{svmrank, ltrGD, GBRank,u-rank}, and listwise \cite{burges2010ranknet, cao2007, softrank, adarank, ListMLE} methods. The pointwise methods such as McRank \cite{mcrank} and PRank \cite{prank} regard ranking as a classification or regression problem and predict an item's relevance score by taking feature representation of one item at a time. The pairwise methods usually model two items simultaneously and convert the ranking to a pairwise classification problem to optimize the relative positions of two items. SVMRank \cite{svmrank} is one of the most famous pairwise learning-to-rank methods built upon the SVM algorithm. 
The listwise methods directly maximize the ranking metrics of lists. For example, LambdaMART \cite{burges2010ranknet} combines the boost tree model MART \cite{sgb,GBM} and the LambdaRank \cite{lambdarank} to optimize NDCG directly.

The experiment section discusses how different types of learning-to-rank methods affect the performance of the reranking models.
\subsection{Reranking}
Compared to ranking methods, reranking methods are mainly multivariate, taking as inputs the whole lists provided by the initial ranker and emphasizing the mutual influence between items. 

Early reranking models mainly focus on modeling the mutual influence of items. According to the mechanism they employ, these methods can be roughly divided into three categories: RNN-based \cite{seq2slate, dlcm, feng2021revisit}, Transformer-based \cite{prm, setrank}, and GNN-based \cite{irgpr} methods. For example, DLCM \cite{dlcm} applies GRU to encode the whole ranking list into the representation of items. Seq2Slate \cite{seq2slate} uses the pointer network with a decoder to generate the reranked list directly. Due to Transformer's ability to model the interaction between any two items, PRM \cite{prm} adopts it to encode the mutual influences between items. PFRN \cite{huang2020personalized} utilizes self-attention to model the context-aware information and conducts personalized flight itinerary reranking. PEAR \cite{pear} employs contextualized transformer to model the item contexts from both the initial list and the historical list. The GNN-based model IRGPR \cite{irgpr} explicitly models item relationships by aggregating relational information from neighborhoods. Besides, GSF \cite{GSF} adopts a DNN architecture to learn multivariate scoring functions by enumerating all feasible permutations of a given size. MIDNN \cite{miDNN} employs global feature extension and DNN to incorporate mutual influences into the feature of an item.

Recently, some work \cite{feng2021revisit, feng2021grn, xi2021contextaware, miDNN} realizes that the reranking operation changes the contexts of each item from the initial list to the reranked list (i.e., the early scoring problem), and they attempt to find the optimal permutation after reranking. PRS \cite{feng2021revisit} first generates the feasible candidate lists with beam search, and then adopts Bi-LSTM to evaluate and select the lists. 
 Reinforcement learning (RL) methods, like GRN \cite{feng2021grn} and EG-Rerank+ \cite{huzhang2020aliexpress}, usually first train an evaluator to predict rewards, and then employ RL methods like PPO to optimize the generator, which generates rankings, with feedback from the evaluator.
URCM \cite{xi2021contextaware} also employs a generator-evaluator solution, where an evaluator with Bi-LSTM and graph attention mechanism is incorporated to estimate the listwise utility via the counterfactual context modeling.
Different from the above generator-evaluator approaches that are usually of\textit{ polynomial inference time}, SetRank \cite{setrank} solves this problem by set modeling. It employs self-attention blocks to get permutation-equivariant representations of items and sort its output to achieve permutation invariant. Setrank aims to find the best permutation directly, considering all items in a ranking list of\textit{ linear inference time}. With permutation-invariant property, the ordering of the initial lists would not affect the output and thus solve the early scoring problem. To avoid complex two-stage (generator-evaluator) solutions, our work takes a similar set modeling approach as SetRank \cite{setrank} by permutation-equivariant function. Moreover, our work considers the dynamic interaction between the user history list and the candidate set, as well as the fine-grained feature-level interaction.

 \section{Problem Formulation}
  \begin{figure*}[t]
    \centering
    \includegraphics[width=0.97\textwidth]{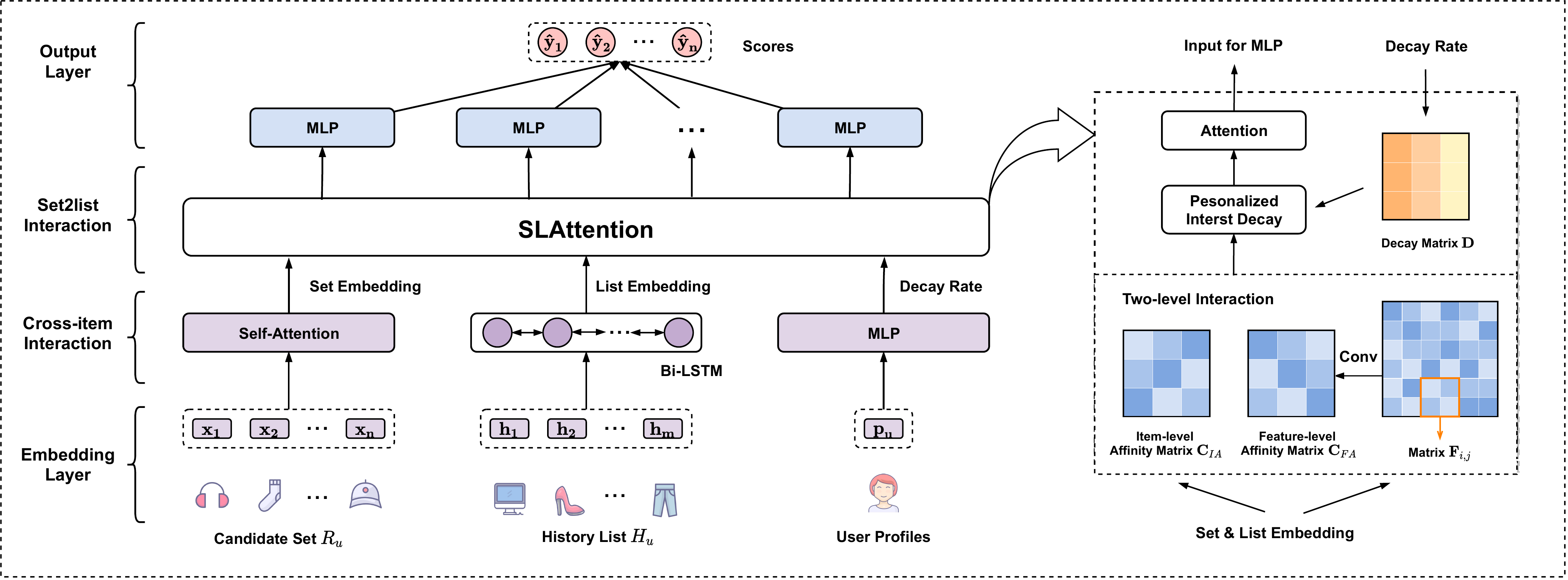}
    \caption{The overall framework of MIR and its sub-modules.}
    \label{fig:framework}
\end{figure*}
    
 A reranking model generally takes as input ordered initial lists arranged by the previous ranker and refines the ranking lists by modeling the mutual influence between items. Given a set of users $\mathcal{U}=\{1,2,...,N_u\}$,
  a set of items $\mathcal{V}=\{1,2,...,N_v\}$,
a history item list $H_u\in\ \Pi_{m}(\mathcal{V})$ with $m$ items that the user $u$ clicked, and an initial ranking 
list $R_u \in \Pi_{n}(\mathcal{V})$ with $n$ items for each user $u\in \mathcal{U}$, 
the reranking model aims to generate a reranking list that maximizes the overall utility and better meets the user's needs. The utility of a list is defined by the expected sum (or weighted sum) of the click probability of each item in the list, e.g., the total number of clicks or the total income (click probability weighted by the bid price). Here, $\Pi_{n}(\mathcal{V})$ is the set of all $n$-permutation of set $\mathcal{V}$. 

 \section{Model Framework}
In this section, we first introduce the overall framework of our proposed MIR and then present the details of each component. The chronological order of user behavior history assists in modeling users' long- and short-term interests, hence we regard the user history as an \textit{ordered list}. Similar to SetRank \cite{setrank}, MIR formulates the candidate items as an \textit{unordered set} to avoid the early-scoring problem \cite{feng2021grn, xi2021contextaware, setrank}. MIR aims to learn a permutation-invariant model by providing a permutation-equivariant function. A permutation-equivariant function permutes the output according to the permutation of the input, and its formal definition is given in Section \ref{Theore}. Therefore, a permutation-invariant ranking model can be achieved by sorting the output of the permutation-equivariant function.

 As depicted in Figure \ref{fig:framework}, MIR consists of an embedding layer, a cross-item interaction module, a set2list interaction module, and an output layer. Firstly, the embedding layer converts categorical features to dense embeddings.
 Then, the cross-item interaction module introduces intra-set and intra-list 
 interaction to model the mutual influences within the candidate set and history list, respectively. Next, in set2list interaction module, SLAttention is designed to capture the dynamic interaction between the candidate set and the history list. Finally, the extracted multi-level interaction information is integrated by the output layer to obtain the final prediction. 

\subsection{Embedding Layer}
MIR takes the items in the initial ranking list $R_u$, the history list $H_u$, and the user profile of the user $u$ as input. We apply the embedding layer on the corresponding sparse raw features to obtain low-dimensional dense embedding vectors. Specifically, for an item $v\in R_u$, its raw features are usually composed of several categorical features (e.g., item\_id, category) and dense features (e.g., price). For the $i$-th categorical feature of item $v$, we project it from a high-dimensional sparse space (one-hot vector) to a lower-dimensional dense space via a learned projection matrix, and get its embedding $\mathbf{e}_{v,i}\in \mathbb{R}^{d_e}$ of size $d_e$. Then, we concatenate all the embeddings for categorical features, together with dense features to generate the embedding vector of an item from the candidate set $\mathbf{x}_v=[\mathbf{e}_{v,1}\oplus...\oplus \mathbf{e}_{v,k}\oplus \mathbf{x}_v^d]\in\mathbb{R}^{d_x}$, where $k$ is the number of the categorical features, $\mathbf{x}^d_v\in \mathbb{R}^{d_\Delta}$ of size $d_\Delta$ contains the dense features of item $v$, $d_x=kd_e+d_\Delta$ is the size of item embedding, and $\oplus$ represents the concatenation operation. Similarly, we can derive the embedding vector $\mathbf{h}_v\in\mathbb{R}^{d_x}$ of size $d_x$ for each item $v$ in history list $H_u$, and the corresponding user embedding vector $\textbf{p}_u\in\mathbb{R}^{d_u}$ of size $d_u$ for user $u$. The projection matrix is shared for items from the candidate set and the history list. 


 \subsection{Cross-item Interaction}
After the embedding layer, we utilize cross-item interaction to model the mutual influence within the history list and candidate set, and obtain \textit{the list embedding} and \textit{the set embedding}, respectively. The mutual influence between items in the ranking list is an essential factor for the reranking. An item's utility is not independent and varies with different contexts, so items should be aware of other items inside the candidate set. Moreover, the sequential dependencies encoded in the history list reflect the characteristics of the user's evolving interests as a whole. Therefore, we introduce intra-set and intra-list interaction before modeling the dynamic interaction between the candidate set and the history list. 

 \textbf{Intra-set Interaction} is designed for capturing the cross-item relationship within the candidate set. As we employ the permutation-equivariant function to achieve permutation-invariance, the intra-set interaction should preserve the permutation-equivariant property to the candidate set. In other words, it should permute its output according to the permutation of the input.
 As such, we adopt the multi-head self-attention mechanism \cite{transformer} to extract the intra-set interaction. This mechanism enables us to model mutual influences between any two items directly and is proved to be permutation-equivalent \cite{setrank}. We stack all the item embedding vectors $\mathbf{x}_i\in\mathbb{R}^{d_x}, i=1,2,...,n$ of the ranking list $R_u$ and obtain a matrix $\mathbf{V}\in\mathbb{R}^{n\times d_x}$, where $d_x$ are the size of item embedding. The self-attention mechanism is defined as
 \begin{equation}
     \mathbf{A}_{cross}=\text{Attention}(\mathbf{V})=\text{softmax}\left(\frac{\mathbf{V}\mathbf{V}^\top}{\sqrt{d_x}}\right)\mathbf{V}\,,
 \end{equation}
 where $\mathbf{A}_{cross}\in \mathbb R^{n\times d_x}$ is the attended matrix, and $\sqrt{d_x}$ is used to stabilize gradients during training. We denote the $i$-th row of matrix $\mathbf{A}_{cross}$ as $\mathbf{a}_i$, the set embedding for item $i$ in ranking list $R_u$.
 
 \textbf{Intra-list Interaction} models the mutual influence between items in the history list. Unlike the candidate set, the history list carries the temporal pattern of the user's interests and preferences, providing important guidance for reranking. To better leverage the temporal pattern in the history list, We use a Bi-LSTM~\cite{schuster1997bidirectional} for modeling users' evolving interests. Bi-LSTM is a lightweight network to handle sequence data, commonly used in previous reranking work~\cite{feng2021grn, feng2021revisit, xi2021contextaware}. Note that other networks like GRUs~\cite{cho2014gru} or self-attention with positional embeddings~\cite{transformer} could also be applied.

Let $\overrightarrow{\mathbf{q}_i}\in\mathbb{R}^{d_h}$ of size $d_h$ be the forward output state of the $i$-th item in Bi-LSTM. That is $\overrightarrow{\mathbf{q}_i}=\overrightarrow{LSTM} ([\mathbf{q}_{i-1}, \mathbf{h}_i, \mathbf{c}_{i-1}])$, where $\overrightarrow{LSTM}(\cdot)$, $\mathbf{q}_{i-1}$, $\mathbf{h}_i$, and $\mathbf{c}_{i-1}$ are the forward LSTM unit, the output of previous item $i-1$, the item embedding of current item $i$, and the cell vector of item $i-1$. Similarly, we can obtain the backward output state $\overleftarrow{\mathbf{q}_i}$. Then, we concatenate $\overrightarrow{\mathbf{q}_i}$ and $\overleftarrow{\mathbf{q}_i}$ to get the list embedding $\mathbf{q}_{i}=[\overrightarrow{\mathbf{q}_i}\oplus \overleftarrow{\mathbf{q}_i}]\in\mathbb{R}^{2d_h}$ of item $i$ in the history list $H_u$ for user $u$.

 \subsection{Set2list Interaction}\label{subsect:set2list}
 After the low-level cross-item interaction within the candidate set (intra-set interaction) and the history list (intra-list interaction), we can model the high-level dynamic interactions between the candidate set and the history list (set2list interaction). 
 However, the interaction is not straightforward since set and list are different data structures, and the number of items and the dimension of features may also be different. Therefore, the set2list interaction should meet several requirements: (i) it must be capable of fusing information from both sources to extract useful information for reranking. (ii) When interacting, it has to consider the asymmetric structure of sets and lists, i.e., it needs to be sensitive to the temporal order of the history list while keeping permutation-equivariant to the candidate set. (iii) It is supposed to be user-specific and consider the personalized impact of the history lists for reranking.

 Inspired by co-attention \cite{co-attention} for Visual Question Answering, we propose a specific attention structure, SLAttention, for set2list interaction. The original co-attention simply adopts a symmetric structure to generate spatial maps, highlighting image regions relevant to answering the question. SLAttention further explores how to deal with the asymmetry between two different data structures and consider the reranking task's characteristics. As illustrated in Figure \ref{fig:framework}, SLAttention firstly explores both item-level and feature-level interactions between the candidate set and history list, and then introduces personalized interest decay to detect the personalized influence of the user history on the reranked items. Finally, an asymmetric attention is utilized to extract information from the candidate set and history list. The rows and columns of the matrices in Figure \ref{fig:framework} represent items in the candidate set and history list, respectively.

\subsubsection{Affinity Matrix} 
 SLAttention attends to the candidate set and history list simultaneously. We fuse the information from the candidate set and history list by calculating the similarity between any pairs of items in set and list. This similarity, i.e., affinity matrix, is aggregated from both item-level and feature-level interactions. 
 
 Item-level interaction 
 takes as inputs the item feature and the embedding obtained in intra-set and intra-list interaction. We denotes representation matrix of candidate set as $\mathbf{S}\in\mathbb{R}^{n\times (2d_x)}$, whose $i$-th row, $\mathbf{s}_i$, is the concatenation of item embedding $\mathbf{x}_i$ and set embedding $\mathbf{a}_i$ of item $i$ in initial ranking list $R_u$. For history list, we denote $\mathbf{L}\in\mathbb{R}^{m\times (d_x+2d_h)}$ as its representation matrix and its $j$-th row, $\mathbf{l}_j$, 
 is the concatenation of item embedding $\mathbf{x}_j$ and list embedding $\mathbf{q}_j$ 
 of item $j$ in history item list $H_u$. Then, the item-level affinity matrix $\mathbf{C}_{IA}\in \mathbb{R}^{n\times m}$ in Figure \ref{fig:framework} is calculated by
  \begin{equation}
     \mathbf{C}_{IA}=\tanh{(\mathbf{S}\mathbf{W}_{IA}\mathbf{L}^T)}\,,
     \label{eq:AI}
 \end{equation}
 where the learnable matrix $\mathbf{W}_{IA}\in\mathbb{R}^{2d_x\times (d_x+2d_h)}$ is the importance of the association between any pair of items in set and list. 
 
 Item-level interaction can leverage the high-order feature interaction between candidate set and history list. However, in recommender systems, items are usually characterized by categorical features, e.g., category, price level. Item-level interaction takes each item as a whole and the category-level semantic information is lost. The interactions among these fine-grained features can provide useful information for reranking. 
 Thus, we propose feature-level interaction to explicitly learn the categorical feature interaction, which only takes as input the embeddings of the categorical features of items. For item $i$ in the candidate set, we denote $\mathbf{E}_S^i\in\mathbb{R}^{k\times d_e}$ as its representation matrix, with the categorical embedding $\mathbf{e}_{i,z}$ as the $z$-th row, and $k$ is the number of the categorical features. Similarly, we can obtain the representation matrix $\mathbf{E}_L^j\in\mathbb{R}^{k\times d_e}$ for any item $j$ in history list $H_u$. Then, the feature-level affinity matrix $\mathbf{C}_{FA} \in\mathbb{R}^{n\times m}$ in Figure \ref{fig:framework} is 
 computed by
  \begin{equation}
  \begin{split}
     \mathbf{F}_{i,j} &=\tanh{(\mathbf{E}_S^i\mathbf{W}_{FA}(\mathbf{E}_L^j)^T)}\, , \\
     \mathbf{C}_{FA}(i,j)&=\sum_{s=1}^k\sum_{t=1}^k \mathbf{F}_{i,j}(s,t)\mathbf{W}_c(s,t)\, , \\
    \end{split}
    \label{eq:FA}
 \end{equation}
 where $\mathbf{W}_{FA}\in\mathbb{R}^{d_e\times d_e}$ denotes the weights for feature interaction and the weights in $\mathbf{W}_c\in\mathbb{R}^{k\times k}$ represent the importance of interaction results. We refer $\mathbf{C}_{FA}(i,j)$ as the element lies in the $i$-th row and $j$-th column of matrix $\mathbf{C}_{FA}$. By performing $\mathbf{F}_{i,j}=\tanh{(\mathbf{E}_S^i\mathbf{W}_{FA}(\mathbf{E}_L^j)^T)}$ for any $i=1,\ldots, n$ and $j=1,\dots, m$, we get $nm$ number of $\textbf{F}$s. Then, weights $\mathbf{W}_c$ and the summation operation are applied to map the matrix $\mathbf{F}_{i,j}\in\mathbb{R}^{k\times k}$ to a scalar $\mathbf{C}_{FA}(i,j)\in\mathbb{R}$. This operation can also be interpreted as convolution, as shown in Figure \ref{fig:framework}. Finally, we get the final affinity matrix $\mathbf{C}_{A}$ by combining the feature-level and the item-level affinity matrices,
  \begin{equation}
  \mathbf{C}_{A}=\mathbf{C}_{IA}+\mathbf{C}_{FA}\,.
 \end{equation}
 

 \subsubsection{Personalized Interest Decay} 
Users' demands and interests are reflected in their historical behavior. Nevertheless, they are not set in stone and may evolve over time. Inspired by the studies in human behaviors and recommendation \cite{Rahimi13locationrecommendation, Tsai2004, CTRec, guidotti2017,Bhagat18}, we consider two kinds of time-sensitive interests: long-term interests and short-term interests. Long-term interests refer to the persistent interests or demands of similar products or types, e.g., daily necessities and stable preferences. In comparison, short-term interests are strongly related to the products browsed or purchased recently, e.g., buying phone cases after buying a phone. 
Thus, personalized interest decay is utilized to capture the complex time-sensitive correlations in history lists and better detect users' current demands.

Inspired by \cite{CTRec,Bhagat18}, we add a multiplicative exponential decay term to detect how users' history lists and the time interval impact the short-time interests. 
Users' short-term interests not only depend on users' recent behaviors but also on the characteristics of the users. Different users may have different ranges of interests. For example, some people's interests may shift quickly, with a narrow window ahead in their history lists being their short-term interests. While others shift their interests very slowly, with a long window of history lists being their short-term interests. Therefore, we use user embedding vector $\mathbf{p}_u$ to learn personalized interest decay rate $\theta_u$ and get the decay vector $\mathbf{d}$ by
   \begin{equation}
     \theta_u =g(\mathbf{p}_u)\, ,\,\,
     \mathbf{d} =e^{-\theta_u \mathbf{t_u}}\, , \\
    \label{eq:decaymat}
 \end{equation}
 where $\theta_u>0$ and $\mathbf{t_u}\in\mathbb{R}^{m}=\{t_1, t_2, ..., t_m\}$ denotes the time interval from when the item in history list was clicked by user $u$ to the present. This time interval can be expanded to general temporal distance measures, like position, to deal with the absence of the timestamp. We use a two-layer feed-forward neural network for function $g(\cdot)$, with the LeakyReLU activation function.
 
We refer to $\mathbf{C}_{A}$ as the long time interest matrix and multiply it by the exponential decay vector to get the short time interest matrix. 
Notes that the decay vector $\mathbf{d}\in\mathbb{R}^{m}$ is only applied to the dimension of history list, so we construct a matrix $\mathbf{D}\in\mathbb{R}^{n\times m}$ in Figure \ref{fig:framework} with each row being $\mathbf{d}$ to maintain its permutation-equivariant to the candidate set. Finally, we have
 \begin{equation}
     \mathbf{C} = \mathbf{C}_{A} + \mathbf{C}_{A}\odot \mathbf{D}\,.
     \label{eq:PID}
 \end{equation}
 
 \subsubsection{Attention}
The personalized affinity matrix $\mathbf{C}$, which combines long-term and short-term correlation between history list and the candidate set, is utilized to predict the attention weights via the following steps,
 \begin{equation}
    \begin{split}
    \mathbf{Q}_S=\tanh{(\mathbf{S}\mathbf{W}_s+\mathbf{C}(\mathbf{L}\mathbf{W}_l))}\, ,\,\,
    \mathbf{Q}_L=\tanh{(\mathbf{S}\mathbf{W}_s\mathbf{C})} \\
    \mathbf{A}_S=\text{softmax}(\mathbf{Q}_S)\, ,\,\,
    \mathbf{A}_L=\text{softmax}(\mathbf{Q}_L)
    \end{split}
    \label{eq:att1}
 \end{equation}
 where $\mathbf{W}_s\in\mathbb{R}^{2d_x\times n },\mathbf{W}_l\in\mathbb{R}^{(d_x+2d_h)\times n}$ are learnable weight matrices. Attention weight matrices $\mathbf{A}_S\in\mathbb{R}^{n\times n}$ and $\mathbf{A}_L\in\mathbb{R}^{n\times m}$ are the attention probabilities of items in the candidate set and history list, respectively, which preserve helpful information for reranking. Since the goal of reranking is to rerank the candidate items, only the influence from the history list to the candidate set benefits the reranking. Therefore, the calculation of $\mathbf{A}_S$ involves both the candidate set and the history, whereas $\mathbf{A}_L$ considers only the candidate items. Based on the above attention weights, the set and list attention vectors are acquired as the weighted sum of items in the candidate set and history list, i.e., 
  \begin{equation}
    \hat{\mathbf{S}}=\mathbf{A}_S\mathbf{S}\, ,\,\, \hat{\mathbf{L}}=\mathbf{A}_L\mathbf{L}\, ,\\
    \label{eq:att2}
 \end{equation}
 where $\hat{\mathbf{S}}=\{\hat{\mathbf{s}}_1, \hat{\mathbf{s}}_2,...,\hat{\mathbf{s}}_n\}$ and $\hat{\mathbf{L}}=\{\hat{\mathbf{l}}_1, \hat{\mathbf{l}}_2,...,\hat{\mathbf{l}}_n\}$ are the interacted representation matrices containing useful information from both candidate set and history list. 
 
 \subsection{Output Layer}
 As a common and powerful technique in reranking, multi-layer perception (MLP) is integrated into the output layer. Hence, taking the concatenation of the user embedding $\mathbf{p}_u$, the item embedding $\mathbf{x}_{v}$, the interacted representation $\hat{\mathbf{s}}_{v}$, and $\hat{\mathbf{l}}_{v}$ as input, the predicted score $\hat{y}_v$ for item $v$ in ranking list $R_u$ can be formalized as follows:
\begin{equation}
\begin{aligned}
     \hat{y}_v=\text{MLP}(\mathbf{p}_{u}\oplus \mathbf{x}_{v}\oplus \hat{\mathbf{s}}_v\oplus\hat{\mathbf{l}}_v;\bm{\Theta})~,
\end{aligned}
\label{eq:g-funct} 
\end{equation}
where LeakyReLU activate function is applied in MLP, $\bm{\Theta}$ denotes the parameters of MLP, and $\oplus$ represents the concatenation operation. 
The final rankings, thus, can be achieved by sorting the items according to the predicted scores.

Given the click label $\mathbf{y}_u=\{y_1,y_2,...,y_n\}$ for ranking list $R_u$, our model can be optimized via binary cross-entropy
loss function, which is defined as follows
\begin{equation}
    \mathcal{L}=\sum_{u\in \mathcal{U}}\sum_{v\in R_u} y_v\log \hat{y}_v +(1-y_v)\log (1-\hat{y}_v)\,.
\end{equation}

\subsection{Theoretical Analysis}\label{Theore}
In this section, we first discuss the permutation-invariant property of MIR, followed by the complexity analysis.

\medskip
\noindent
\textbf{Permutation-invariance.}
Our goal is to construct a permutation-invariant reranking model, which means that any permutation of the items would not change the output ranking. In this work, we rely on a permutation-equivariant function, defined in Definition~\ref{def:perm-inv}, to achieve a permutation-invariant model. 
\newtheorem{defi}{Definition}

\begin{defi}
Let $\Pi_N$ be the set of all permutations of indices $\{1,2,...,N\}$, a function $f:X^N\to Y^N$ is permutation equivariant iff for any permutation $\pi\in\Pi_N$, 
$$f([x_{\pi(1)}, \ldots, x_{\pi(N)}])=[f(X)|_{\pi(1)},\ldots,f(X)|_{\pi(N)}]\,,$$
where $X = [x_1,\ldots, x_N ]$ is a set with the order of $\{1, \ldots, N\}$, and $f(X)|_{\pi(i)}$ is the $\pi(i)$-th dimension of $f(X)$.
\label{def:perm-inv} 
\end{defi}

Note that the parameters in the embedding layer and output layers are all element-wise, which do not affect the permutation-equivariant property. Self-attention block, used in intra-set interaction, is proved to be permutation-equivariant in SetRank \cite{setrank}. We also show that SLAttention structure presented in Section \ref{subsect:set2list} is a permutation-equivariant function, proved in Appendix \ref{appe}.
\newtheorem{propo}{Proposition}
\begin{propo}
SLAttention structure is permutation-equivariant to candidate set.
\label{pro:slattention}
\end{propo}
Previous work \cite{deepSets} shows that the composition of permutation-equivariant functions is also permutation-equivariant. Combining the above proposition and discussions, we obtain that our designed network structure is a permutation-equivariant function.

Built upon a permutation-equivariant function,  a permutation-invariant model (permutation of the input does not affect the output) can be directly derived \cite{setrank}. It is because for different permutations of the same item set,  the output of the permutation-equivariant function permutes according to the way we permute the input -- meaning the output value (reranking score) for each item remains the same. Then, sorting the items according to the reranking score for each item yields the same reranking list. Therefore, we conclude that MIR is permutation-invariant to the candidate set. 

\medskip
\noindent
\textbf{Complexity.}
We also analyze the efficiency of the MIR model.
Considering that the self-attention used in intra-set interaction can be reduced to $O(m)$, the time complexity of cross-item interaction is $O(n+m)$, where $n$ and $m$ are the lengths of the candidate set and history list, respectively. The set2list interaction involves the matrix computation between the set and list, and its cost is $O(mn)$. Therefore, the overall time complexity of MIR is $O(mn)$.

 \section{Experiments}
This section first compares our proposed MIR with the state-of-the-art reranking algorithms on two public datasets and a proprietary industrial dataset. Then, we investigate the impact of several vital components and hyper-parameters, followed by a case study to show how user history dynamically interacts with candidate items.
 \subsection{Experimental Setup}
 \subsubsection{Datasets.}
 We conduct experiments on two public benchmark datasets, including E-commerce Reranking dataset\footnote{https://github.com/rank2rec/rerank} and Ad dataset\footnote{https://tianchi.aliyun.com/dataset/dataDetail?dataId=56}, and a proprietary dataset from a real-world App Store. 
\begin{itemize}
    \item \textbf{PRM Public} contains 
    743,720 users, 7,246,323 items, and 14,350,968 
    records from a real-world e-commerce RS. Each record is a recommendation list consisting of 3 user profile features, 5 categorical, and 19 dense item features. For each user, we use the last one in her interacted lists for reranking, and the positively interacted items in previous ones are used to construct the history lists. 
    \item \textbf{Ad} records 1,140,000 users and 26 million ad display/click logs, with 9 user profiles (e.g., id, age, and occupation), 6 item features (e.g., id, campaign, and brand), and user shopping history of seven hundred million records. Following \cite{feng2021grn, feng2021revisit}, we transform records of each user into ranking lists according to the timestamp of the user browsing the advertisement. Items that have been interacted within five minutes are sliced into a list. 

    \item \textbf{App Store} is collected from a mainstream commercial App Store, from June 2, 2021 to July 1, 2021. The dataset contains 33,646,680 requests and 111,063 Apps. Each App has 32 features, e.g., App developer. Its records are in the form of reranking lists, and each user has a history of behavior, like clicks and downloads.
    
\end{itemize}

 \subsubsection{Initial ranker and baselines.}
 We select three widely adopted ranking algorithms, including DIN, SVMRank, and LambdaMART, to generate initial lists. Those three algorithms use pointwise, pairwise, and listwise loss, respectively. \textbf{DIN} \cite{zhou18kdd} designs a local activation unit to adaptively learn the representation of user interests from history lists to help the ranking. \textbf{SVMRank} \cite{svmrank} is a classic pairwise learning-to-rank model built upon the SVM algorithm. \textbf{LambdaMART} \cite{burges2010ranknet} is a state-of-the-art listwise learning-to-rank algorithm, which optimizes NDCG directly.

 We compare the proposed model with the following state-of-the-art reranking models, listed as follows \footnote{We did not include the results of the generator-evaluator methods, Eg-rerank+ \cite{huzhang2020aliexpress} and URCM \cite{xi2021contextaware}, because with their released code and careful parameter tuning, some of the metrics still worsen than the initial ranking. Possible reasons are: (i) the generator-evaluator models rely on dense and instant feedback labels for effective training, which are originally provided by a simulation environment in their work. While our data is extremely sparse. (ii) the performance of the generator heavily depends on the quality of the evaluator, while it is hard to measure and select a good evaluator -- a minor change in the evaluator leads to a large performance difference of the generator.}.
 \begin{itemize}
     \item \textbf{MIDNN} \cite{miDNN} extracts mutual influence between items in the input ranking list with global feature extension.
     \item \textbf{DLCM} \cite{dlcm} first applies GRU to encode and rerank the top results.
     \item \textbf{GSF} \cite{GSF} uses DNN to learn multivariate scoring functions, where the scores are determined jointly by multiple items in the list.
     \item \textbf{PRM} \cite{prm} employs self-attention to model the mutual influence between any pair of items and users' preferences.
     \item \textbf{SetRank} \cite{setrank} learns permutation-equivariant representations for the inputted items via self-attention.

 \end{itemize}

\begin{table*}[]
\scriptsize
    \caption{Overall performance on benchmark datasets.}
    \centering
    \scalebox{1.05}{
    \setlength{\tabcolsep}{1.0mm}{
\begin{tabular}{cccccc|cccc|cccc|cccc}
\toprule
\multirow{4 }{*}{Ranker} & \multirow{4}{*}{Reranker} & \multicolumn{8}{c|}{PRM Public} & \multicolumn{8}{c}{Ad} \\
\cmidrule{3-18}
 &  & \multicolumn{4}{c|}{@10} & \multicolumn{4}{c|}{@20} & \multicolumn{4}{c|}{@5} & \multicolumn{4}{c}{@10}\\
 \cmidrule{3-18}
 &  & MAP & NDCG & deNDCG & Utility & MAP & NDCG & deNDCG & Utility & MAP & NDCG & deNDCG & Utility & MAP & NDCG & deNDCG & Utility \\
 \midrule
\multirow{7}{*}{DIN} & initial & 0.1929 & 0.2290 & 0.2298 & 1.2289 & 0.1976 & 0.3318 & 0.3324 & 1.7986 & 0.5930 & 0.6660 & 0.6654 & 2.2416 & 0.6028 & 0.6941 & 0.6940 & 2.3574 \\
 & MIDNN & 0.2986 & 0.3399 & 0.3124 & 1.3267 & 0.2907 & 0.4200 & 0.3965 & 1.8418 & 0.5991 & 0.6705 & 0.6710 & 2.2174 & 0.6093 & 0.6990 & 0.6994 & 2.3345 \\
 & DLCM & 0.3002 & 0.3422 & 0.3146 & 1.3431 & 0.2919 & 0.4227 & 0.3991 & 1.8426 & 0.5998 & 0.6715 & 0.6715 & 2.3126 & 0.6094 & 0.6992 & 0.6995 & 2.4257  \\
 & GSF & 0.2989 & 0.3402 & 0.3127 & 1.3283 & 0.2909 & 0.4199 & 0.3964 & 1.8418 & 0.5995 & 0.6710 & 0.6713 & 2.2341 & 0.6097 & 0.6993 & 0.6996 & 2.3516  \\
 & PRM & 0.3026 & 0.3446 & 0.3161 & 1.3423 & 0.2940 & 0.4252 & 0.4011 & 1.8653 & 0.6014 & 0.6722 & 0.6725 & 2.2350 & 0.6117 & 0.7006 & 0.7011 & 2.3493   \\
 & SetRank &0.3003 & 0.3413 & 0.3118 & 1.3192 & 0.2919 & 0.4207 & 0.3951 & 1.8320 & 0.6007 & 0.6718 & 0.6719 & 2.2457 & 0.6101 & 0.6995 & 0.6997 & 2.3624  \\
 & \textbf{MIR} &  \textbf{0.3087*} & \textbf{0.3511*} & \textbf{0.3239*} & \textbf{1.3906*} & \textbf{0.2989*} & \textbf{0.4310*} & \textbf{0.4078*} & \textbf{1.9064*} & \textbf{0.6068*} & \textbf{0.6768*} & \textbf{0.6771*} & \textbf{2.3807*} & \textbf{0.6164*} & \textbf{0.7044} & \textbf{0.7048*} & \textbf{2.4918*} \\

 \midrule
\multirow{7}{*}{SVMRank} & initial & 0.1746 & 0.2057 & 0.2093 & 1.1572 & 0.1815 & 0.3079 & 0.3110 & 1.7176 & 0.5864 & 0.6607 & 0.6603 & 2.1978 & 0.5964 & 0.6889 & 0.6888 & 2.3142 \\
 & MIDNN &  0.2982 & 0.3394 & 0.3113 & 1.3276 & 0.2905 & 0.4193 & 0.3948 & 1.8409 & 0.5975 & 0.6694 & 0.6697 & 2.2192 & 0.6074 & 0.6972 & 0.6975 & 2.3353 \\
 & DLCM &  0.2975 & 0.3383 & 0.3094 & 1.3120 & 0.2896 & 0.4185 & 0.3933 & 1.8293 & 0.5991 & 0.6708 & 0.6712 & 2.3236 & 0.6090 & 0.6983 & 0.6987 & 2.4157 \\
 & GSF & 0.2990 & 0.3404 & 0.3120 & 1.3287 & 0.2910 & 0.4200 & 0.3952 & 1.8417 & 0.5987 & 0.6702 & 0.6704 & 2.2354 & 0.6085 & 0.6980 & 0.6983 & 2.3486 \\
 & PRM & 0.3005 & 0.3414 & 0.3116 & 1.3175 & 0.2919 & 0.4210 & 0.3951 & 1.8328 & 0.5997 & 0.6705 & 0.6705 & 2.1679 & 0.6098 & 0.6988 & 0.6990 & 2.2842 \\
 & SetRank & 0.3002 & 0.3418 & 0.3120 & 1.3211 & 0.2920 & 0.4209 & 0.3949 & 1.8320 & 0.5980 & 0.6698 & 0.6701 & 2.3118 & 0.6079 & 0.6975 & 0.6979 & 2.4237 \\
 & \textbf{MIR} & \textbf{0.3084*} & \textbf{0.3514*} & \textbf{0.3230*} & \textbf{1.3866*} & \textbf{0.2993*} & \textbf{0.4308*} & \textbf{0.4059*} & \textbf{1.8989*} &  \textbf{0.6056*} & \textbf{0.6760*} & \textbf{0.6765*} & \textbf{2.3683*} & \textbf{0.6151*} & \textbf{0.7029} & \textbf{0.7033*} & \textbf{2.4776*} \\
 \midrule
\multirow{7}{*}{LambdaMART} & initial &  0.1820 & 0.2139 & 0.2158 & 1.1569 & 0.1879 & 0.3155 & 0.3174 & 1.7188 & 0.5897 & 0.6633 & 0.6629 & 2.1783 & 0.5997 & 0.6915 & 0.6915 & 2.2948 \\
 & MIDNN & 0.2984 & 0.3396 & 0.3127 & 1.3269 & 0.2906 & 0.4196 & 0.3967 & 1.8427 & 0.5979 & 0.6697 & 0.6704 & 2.2329 & 0.6077 & 0.6975 & 0.6980 & 2.3481 \\
 & DLCM & 0.2984 & 0.3394 & 0.3118 & 1.3149 & 0.2906 & 0.4190 & 0.3954 & 1.8295 & 0.5995 & 0.6710 & 0.6712 & 2.2801 & 0.6093 & 0.6988 & 0.6990 & 2.3739 \\
 & GSF &  0.2988 & 0.3400 & 0.3130 & 1.3293 & 0.2909 & 0.4200 & 0.3969 & 1.8441 & 0.5991 & 0.6706 & 0.6710 & 2.2735 & 0.6092 & 0.6986 & 0.6990 & 2.3873 \\
 & PRM & 0.3002 & 0.3415 & 0.3129 & 1.3156 & 0.2919 & 0.4210 & 0.3966 & 1.8299 & 0.6004 & 0.6714 & 0.6712 & 2.2171 & 0.6107 & 0.6996 & 0.6997 & 2.3327 \\
 & SetRank &0.2999 & 0.3413 & 0.3132 & 1.3210 & 0.2917 & 0.4206 & 0.3966 & 1.8333 & 0.6001 & 0.6716 & 0.6715 & 2.2789 & 0.6098 & 0.6991 & 0.6993 & 2.3923 \\
 &\textbf{MIR} &\textbf{0.3083*} & \textbf{0.3511*} & \textbf{0.3247*} & \textbf{1.3907*} & \textbf{0.2991*} & \textbf{0.4301*} & \textbf{0.4073*} & \textbf{1.8998*} & \textbf{0.6060*} & \textbf{0.6762*} & \textbf{0.6765*} & \textbf{2.3685*} & \textbf{0.6157*} & \textbf{0.7037} & \textbf{0.7042*} & \textbf{2.4799*}\\

 \bottomrule
\end{tabular}
}
}
\footnotesize \flushleft\hspace{0cm} $*$ denotes statistically significant improvement (measured by t-test with $p$-value $<$ 0.05) over the best baseline.
    \label{tab:overall}
\end{table*}
 
 \subsubsection{Evaluation metrics}
 
 Our proposed model and baselines are evaluated by both ranking and utility metrics. For ranking metrics, we adopt the widely-used \textit{MAP@K} and \textit{NDCG@K} \cite{ndcg} following previous work \cite{prm,feng2021revisit,feng2021grn}. Nevertheless, \textit{MAP@K} and \textit{nDCG@K} are computed with click labels in the log data, which may be biased to certain recommendation scenarios. So we also provide \textit{deNDCG@K}, an evaluation metric debiased by inverse propensity score (IPS) weighting, according to \cite{dendcg}. We estimate the propensity score following \cite{u-rank}.
For the utility metrics, the \textit{Utility@K} for the public datasets PRM Public and Ad is the debiased expected number of clicks, which is also debiased by IPS weighting. For the proprietary dataset App Store, as the objective of the platform is to maximize the total revenue, the  \textit{Utility@K} is the expected revenue. The detailed description can be found in Appendix \ref{metrics}.

The PRM Public dataset provides a fixed length of reranking lists of 30, so we set $K=10,20$. For the Ad dataset, we divide the lists by timestamps, and their lengths are relatively short, so we conduct reranking only at the top-$10$ items given by the initial ranker and set $K=5,10$. For the App Store dataset, since the maximum number of positions is set to 20, we set $K=5,10$. 
 


 \subsubsection{Reproducibility.} 
 The implementation of our proposed MIR is publicly available \footnote{Our code is available at https://github.com/YunjiaXi/Multi-Level-Interaction-Reranking}. We implement our model and baselines with Adam \cite{adam} as optimizer. We use the last 30 items clicked by the user as their history. According to different scenarios, the maximum length of initial lists is set to 30, 10, and 20 for PRM Public, Ad, and App Store, respectively. The learning rate is selected from $\{2\times 10^{-5}, 3\times 10^{-5}, 1\times 10^{-4}, 2 \times 10^{-4}\}$ and the parameter of L2-Regularization from $\{1\times 10^{-5}, 2\times 10^{-5},3\times 10^{-5}, 5\times 10^{-5}\}$. The batch size and hidden size are set to 16 and 64. The embedding size of the categorical feature is set to 16, and the architecture of MLP is set to [500, 200, 80]. To ensure a fair comparison, we also fine-tune all baselines to achieve their best performance. 
 

\begin{table}[]
\newcommand{\tabincell}[2]{\begin{tabular}{@{}#1@{}}#2\end{tabular}}
\tiny
    \caption{Overall performance on App Store datasets.}
    \centering
    \scalebox{1.20}{
    \setlength{\tabcolsep}{0.75mm}{
\begin{tabular}{ccccc|cccc}
\toprule
\multirow{2}{*}{Model} & 
  \multicolumn{4}{c|}{@5} & \multicolumn{4}{c}{@10} \\
 \cmidrule{2-9}
 & MAP & NDCG & deNDCG & Utility & MAP & NDCG & deNDCG & Utility  \\
 \midrule
init         & 0.1855           & 0.3549           & 0.3474           & 2.4671           & 0.1809           & 0.4260           & 0.4200           & 3.3646           \\
MIDNN        & 0.2352           & 0.4349           & 0.4340           & 3.5379           & 0.2293           & 0.5014           & 0.5008           & 4.6408           \\
DLCM         & 0.3205           & 0.5074           & 0.5112           & 3.9615           & 0.3145           & 0.5588           & 0.5623           & 4.8690           \\
GSF          & 0.2271           & 0.4253           & 0.4249           & 3.4690           & 0.2213           & 0.4941           & 0.4942           & 4.6046           \\
PRM          & 0.3281           & 0.5132           & 0.5166           & 3.9945           & 0.3222           & 0.5662           & 0.5685           & 4.9185           \\
SetRank      & 0.2591           & 0.4537           & 0.4569           & 3.6234           & 0.2533           & 0.5168           & 0.5194           & 4.6845           \\
\textbf{MIR} & \textbf{0.3449*} & \textbf{0.5301*} & \textbf{0.5337*} & \textbf{4.0964*} & \textbf{0.3396*} & \textbf{0.5815*} & \textbf{0.5838*} & \textbf{5.0014*}  \\
 \bottomrule
\end{tabular}
}
}
\footnotesize \flushleft\hspace{0cm} $*$ denotes statistically significant improvement (measured by t-test with $p$-value $<$ 0.05) over the best baseline.
    \label{tab:appstore}
\end{table}

\begin{table*}[]
\newcommand{\tabincell}[2]{\begin{tabular}{@{}#1@{}}#2\end{tabular}}
\scriptsize
    \caption{Comparison of MIR and its variants on two benchmark datasets.}
    \centering
    \scalebox{1.2}{
    \setlength{\tabcolsep}{0.85mm}{
\begin{tabular}{ccccc|cccc|cccc|cccc}
\toprule
\multirow{4}{*}{Model} & \multicolumn{8}{c|}{PRM Public} & \multicolumn{8}{c}{Ad} \\
\cmidrule{2-17}
 & \multicolumn{4}{c|}{@10} & \multicolumn{4}{c|}{@20} & \multicolumn{4}{c|}{@5} & \multicolumn{4}{c}{@10} \\
 \cmidrule{2-17}
 & MAP & NDCG & deNDCG & Utility & MAP & NDCG & deNDCG & Utility & MAP & NDCG & deNDCG & Utility & MAP & NDCG & deNDCG & Utility \\
 \midrule
MIR-fi & 0.3072 & 0.3503 & 0.3228 & 1.3836 & 0.2983 & 0.4304 & 0.4069 & 1.9011 & 0.6030 & 0.6737 & 0.6738 & 2.2957 & 0.6127 & 0.7015 & 0.7019 & 2.4099 \\
MIR-ii & 0.3067 & 0.3500 & 0.3226 & 1.3834 & 0.2975 & 0.4295 & 0.4062 & 1.8978 & 0.6040 & 0.6745 & 0.6748 & 2.3181 & 0.6136 & 0.7022 & 0.7026 & 2.4307 \\
MIR-dcy & 0.3060 & 0.3490 & 0.3213 & 1.3733 & 0.2970 & 0.4285 & 0.4049 & 1.8878 & 0.6029 & 0.6734 & 0.6739 & 2.2831 & 0.6128 & 0.7016 & 0.7021 & 2.3976 \\
 MIR-SLA & 0.3026 & 0.3447 & 0.3162 & 1.3406 & 0.2945 & 0.4246 & 0.4002 & 1.8553 & 0.6009 & 0.6721 & 0.6726 & 2.2713 & 0.6106 & 0.7000 & 0.7005 & 2.3854 \\
\midrule
MIR-set & 0.3071 & 0.3497 & 0.3224 & 1.3828 & 0.2979 & 0.4294 & 0.4061 & 1.8968 & 0.6034 & 0.6742 & 0.6745 & 2.3231 & 0.6130 & 0.7018 & 0.7021 & 2.4351 \\
MIR-lst & 0.3073 & 0.3508 & 0.3236 & 1.3879 & 0.2985 & 0.4304 & 0.4072 & 1.9020 & 0.6038 & 0.6744 & 0.6745 & 2.3486 & 0.6136 & 0.7022 & 0.7026 & 2.4601 \\
\midrule
MIR-hst & 0.3054 & 0.3487 & 0.3221 & 1.3879 & 0.2967 & 0.4283 & 0.4056 & 1.9000 & 0.6044 & 0.6749 & 0.6754 & 2.3783 & 0.6143 & 0.7028 & 0.7033 & 2.4885 \\
 \midrule
\textbf{MIR} & \textbf{0.3087} & \textbf{0.3511} & \textbf{0.3239} & \textbf{1.3906} & \textbf{0.2989} & \textbf{0.4310} & \textbf{0.4078} & \textbf{1.9064} & \textbf{0.6068} & \textbf{0.6768} & \textbf{0.6771} & \textbf{2.3807} & \textbf{0.6164} & \textbf{0.7044} & \textbf{0.7048} & \textbf{2.4918}\\
 \bottomrule
\end{tabular}
}
}
    \label{tab:ablation}
\end{table*}
 \subsection{Overall Performance}
 \subsubsection{Benchmark datasets.}
 The overall performance on the two benchmark datasets, PRM Public and Ad, is reported in Table \ref{tab:overall}, from which we have several important observations.
 
 First, our proposed MIR significantly and consistently outperforms the state-of-the-art approaches in all metrics under three initial rankers on both datasets. As presented in Table \ref{tab:overall}, MIR performs best with respect to ranking-based metrics \textit{MAP} and \textit{NDCG}, debiasd metric \textit{deNDCG}, and utility-based metric \textit{Utility}. For instance, MIR surpasses the strongest baseline PRM by 2.02\% in \textit{MAP@10}, 2.47\% in \textit{deNDCG@10}, and 3.60\% in \textit{Utility@10} on PRM Public, with DIN as initial ranker. On Ad dataset, MIR also achieves 0.90\%, 0.67\%, and 6.52\% improvement over baseline PRM in \textit{MAP@5}, \textit{deNDCG@5}, and \textit{Utility@5} with DIN. This demonstrates the effectiveness of leveraging the interactions between the candidate set and the users' history with long-short personalized interests in reranking. 

 
  Second, the performance of MIR is stable with respect to different initial rankers on PRM Public dataset. Because the PRM Public dataset provides a fixed length of reranking lists of 30, we do not need padding or cutting the lists to the same length. This indicates that the candidate items are the same under the three initial rankers. MIR achieves a similar performance on the three initial rankings, showing the permutation-invariance property of MIR. For the Ad dataset, we divide the lists by timestamps and their lengths span a wide range, so we conduct reranking only at the top items given by the initial ranker. In other words, the candidate sets produced by different initial rankers are different, leading to the different performance of MIR. We observe that given better candidate sets, the performance of MIR is generally better. Likewise, SetRank shows a similar trend as MIR since it is also insensitive to permutations.
 

 Third, there exists a trade-off between ranking-based and utility-based metrics. Considering the initial ranker, LambdaMART outperforms SVMRank in the ranking-based metrics, yet underperforms SVMRank in the utility-based metric on most occasions. This is also the case for most of the reranking methods. For example, PRM excels in ranking-based metrics, whereas SetRank is superior in boosting utility. Even for the same model, SetRank, it has the opposite performance on \textit{Utility} and \textit{MAP} on Ad under the various initial rankers. Nevertheless, our proposed MIR reaches a balance and delivers better performance in all types of metrics. 
 \subsubsection{Proprietary dataset.}
 We further conduct experiments on a mainstream industrial app store and evaluate MIR directly by real-world
click-through ranking lists.
As presented in Table \ref{tab:appstore}, similar observations as on the public datasets can be observed for the App Store dataset. The improvement of MIR on App Store dataset is more significant than that on public datasets. For example, MIR surpasses the strongest baseline PRM by 5.12\% in \textit{MAP@5}, 3.29\% in \textit{NDCG@5}, 3.31\% in \textit{deNDCG@5}, and 2.55\% in \textit{Utility@5}. 
Such greater improvements may result from the higher quality of the history list, which is collected in real-time. This also demonstrates the effectiveness of MIR and the importance of introducing dynamic interaction between the history list and the candidate set. 


 \subsection{In-depth Analysis}

\subsubsection{Ablation study. }
Several variants of MIR are designed to investigate the effectiveness of the components in MIR, and we conduct a series of experiments on PRM Public and Ad datasets. Firstly, we remove two kinds of cross-item interactions. \textbf{MIR-lst} removes intra-list interaction. \textbf{MIR-set} removes intra-set interaction.
Then, we remove key components from SLAttention. \textbf{MIR-fi} removes feature-level interaction. \textbf{MIR-ii} removes item-level interaction. \textbf{MIR-dcy} removes personalized interest decay. \textbf{MIR-SLA} replaces SLAttention with a simple variant of self-attention \cite{yu2019mcan}, where the candidate items serve as keys and values and items in user history act as queries.
Lastly, the SLAttention structure relies on user history, but there are cases where the user has no history, such as lost or cold start. We also devise a variant to handle these situations. \textbf{MIR-hst} replaces the original history with the history of a similar user at the inference stage. The similarity is computed by the distance between the user profiles.

The comparison of the above variants and original MIR on PRM Public and Ad datasets is shown in Table \ref{tab:ablation}.
After removing each component, the performance has declined to a certain extent w.r.t. all metrics, which demonstrates the effectiveness of each component. 
Compared to the other variants, the performance of the MIR-SLA drops and is even worse than that of PRM in Table \ref{tab:overall}. This shows that the simple employment of self-attention is insufficient to explore the impacts of history on reranking, illustrating the necessity of devising a specific structure for set2list interaction in reranking.
MIR-dcy has roughly the second-worst performance, which reveals the importance of personalized interest decay. The results of MIR-set and MIR-lst illustrate the effectiveness of cross-item interaction before set2list interaction. MIR-hst achieves much better results than baselines on both datasets. This indicates that even when cold-start users exist, good results can still be achieved if similar users' history is available.

The difference in datasets leads to slightly different performances for these variants on the two datasets. Item-level interaction uses the whole item feature and embedding obtained cross-item interaction, while feature-level interaction only covers categorical features. Therefore, for the PRM Public dataset containing many dense features, removing item-level interaction, e.g., MIR-ii, depresses the performance more, compared with removing feature-level interaction. In contrast, the Ad dataset only has categorical features and thus relies more on feature-level interaction, yielding the opposite trend.
The difference between the performance of MIR-hst on the two datasets is also caused by datasets. We use the similarity of user profiles to choose a similar user. There are 3 and 9 features for user profiles on PRM Public and Ad datasets, respectively. Therefore, the user history lists selected on Ad are closer to that of the original user, so that MIR-hist shows better performance on Ad.

\subsubsection{Hyper-parameter Study}
\begin{figure}[h]
    \centering
    \includegraphics[width=0.92\columnwidth]{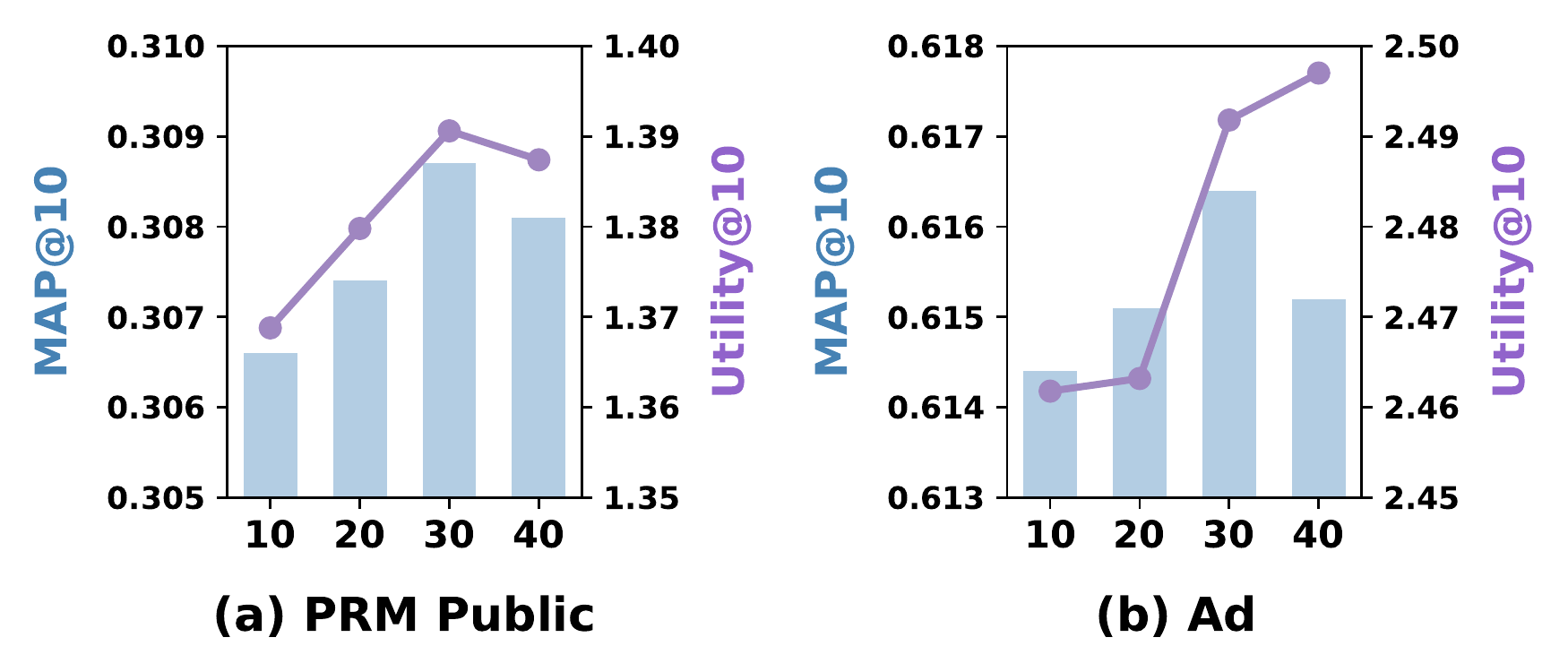}
    \caption{The impact of the length of the user history. }
    \label{fig:history}
    
\end{figure}

Since we focus on leveraging history to support reranking, the length of history is an important hyper-parameter that influences the final results. Thus, we conduct grid-search experiments on public datasets to get a comprehensive understanding of how the length of history affects MIR's performance. With DIN as the initial ranker, We fix all the other hyper-parameters and tune the length of history. Then, we visualize the change of a ranking-based metric \textit{MAP@10} and a utility-based metric \textit{Utility@10} in Figure \ref{fig:history}. We observe both \textit{MAP@10} and \textit{Utility@10} improve sharply from 10 to 30 and then become stable from 30 to 40 on PRM Public and Ad datasets. Although a 40-length history may bring some improvement, it is not cost-effective compared to the additional training time and space required, so we set the length to 30 in our experiments.

\subsubsection{Case study. }
\begin{figure}[h]
    \centering
	\includegraphics[width=0.21\textwidth]{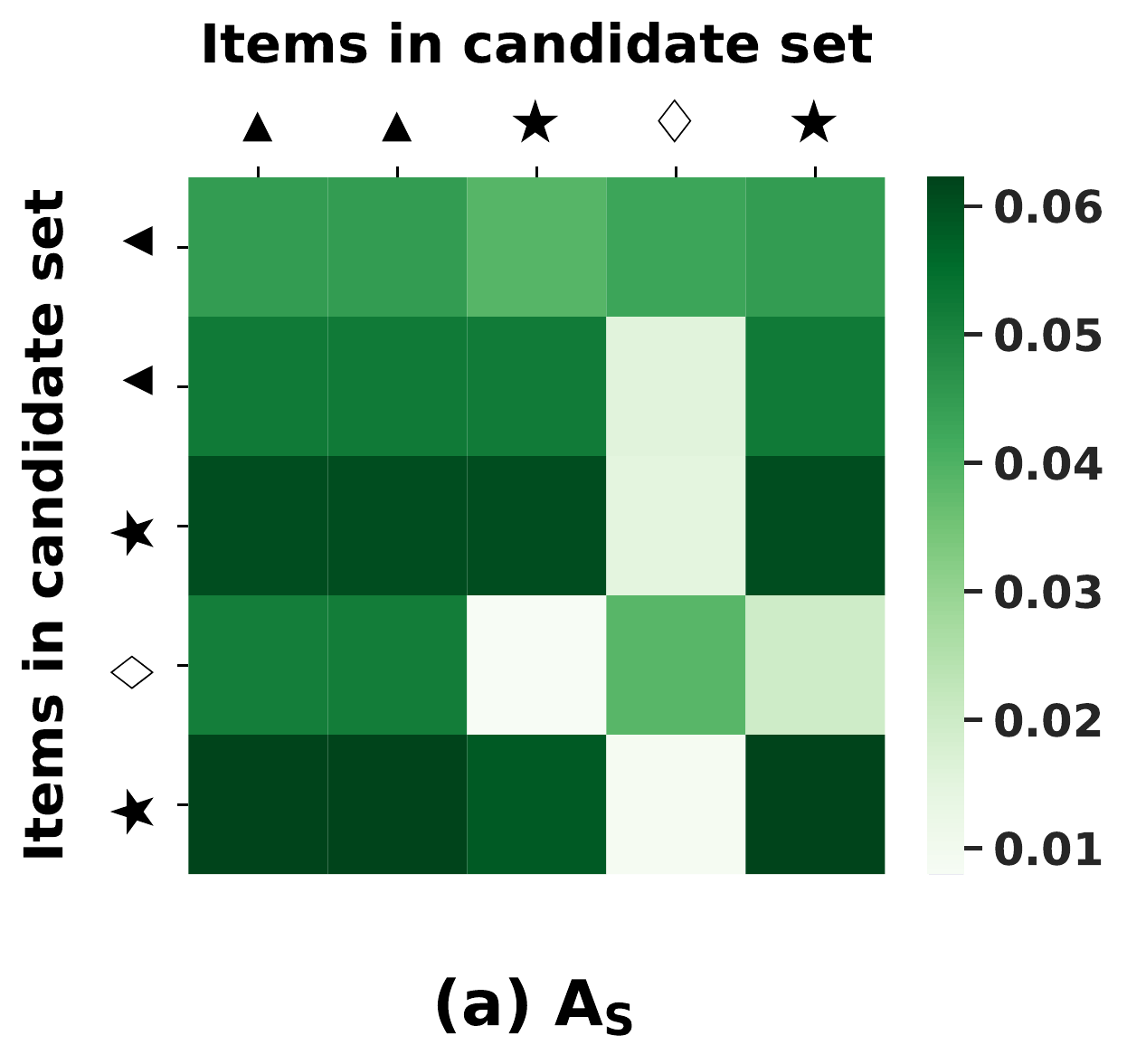}
	\quad
	\includegraphics[width=0.21\textwidth]{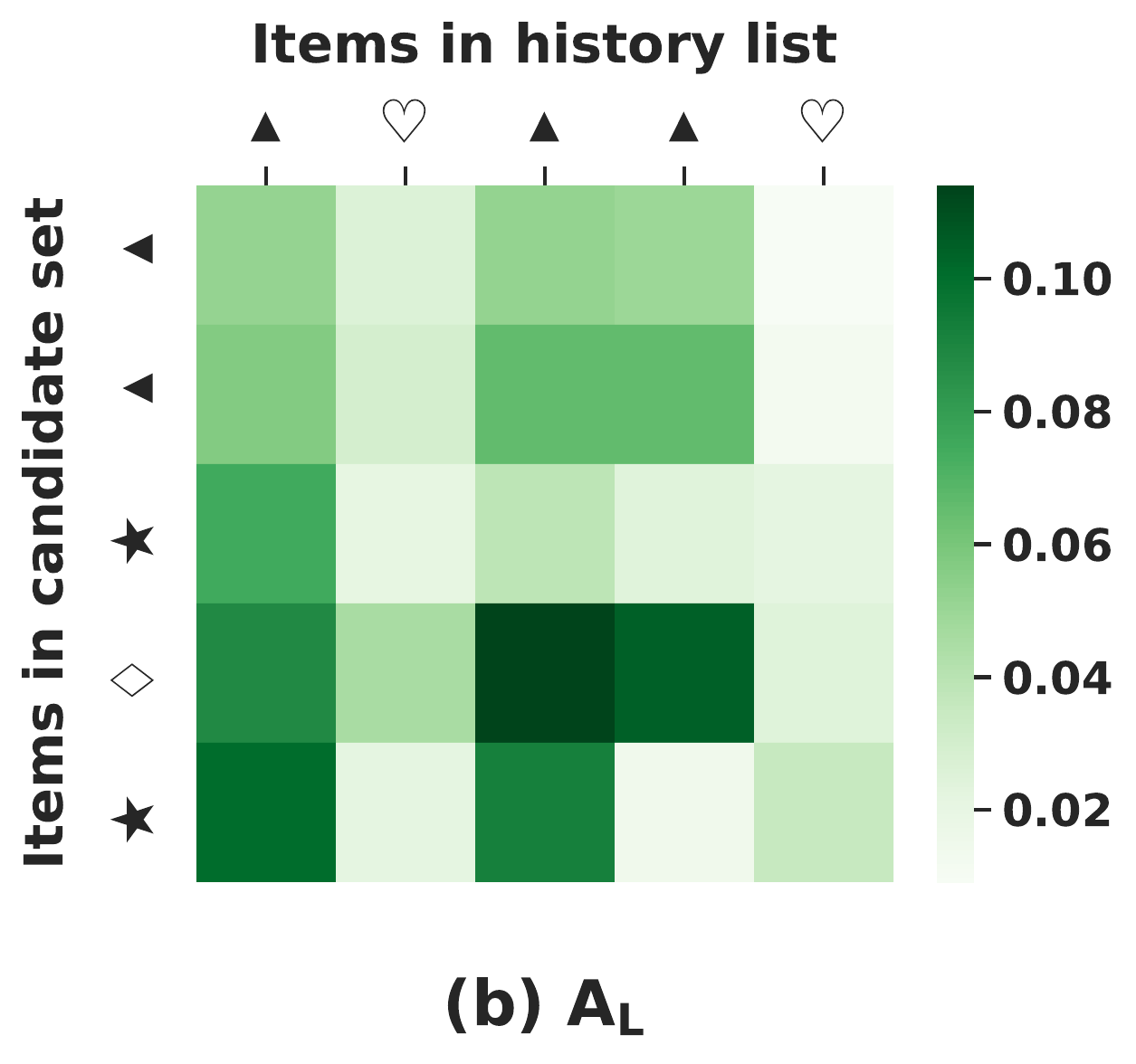}
    \caption{Visualization of the weights of SLAttention.}
    \label{fig:case}
\end{figure}

 To show how SLAttention structure extracts information from the set2list interactions, we visualize the attention coefficients $A_S$ and $A_L$ obtained in Eq. \eqref{eq:att1}. We select a record on PRM Public dataset and take the top five items from the candidate set and history list, respectively. Then, we plot the heatmaps in Figure \ref{fig:case}, with each row being the attention weights of an item on the horizontal axis items. Since the PRM does not provide a specific name for each category, we use symbols to replace the category IDs in horizontal and vertical axes.
 
 In Figure \ref{fig:case}(a), most of the items have lower attention weights for the diamond-shaped items, suggesting that the other items in the candidate set possibly depend less on the diamond item for reranking. The weights of the same type of items are similar in most cases, such as triangular and star-shaped items. From Figure \ref{fig:case}(b) we can see a consistent dependency relationship with (a). The diamond-shaped item shows a similar tendency to be more dependent on the triangular items in history. Items in the candidate set generally have lower weights for heart-shaped items and higher weights for triangular-shaped items. The weights for items of the same type are also more consistent but not identical. This is because we have only used the category here, but other features in the dataset may cause differences. In summary, the SLAttention structure is capable of extracting informative patterns in the set2list interaction.
 
 \section{Conclusion}
 In this work, we address the limitations of previous work in reranking, especially neglecting users' dynamic and
personalized interests from users' history. Thus, we propose a novel end-to-end model, MIR, where the candidate items to be reranked and users' behavior history are formulated as a set and a list, respectively. MIR consists of lower-level cross-item interaction within the candidate set or the history list, and higher-level set2list interaction between them. Moreover, feature-level interactions are incorporated to capture the fine-grained influence. Specifically, we design a SLAttention structure for capturing the set2list interactions, and theoretically analyze its permutation-equivariant property. Extensive experiments show that MIR significantly outperforms the state-of-the-art baselines regarding ranking and utility metrics. 

\begin{acks}
The SJTU team is supported by ``New Generation of AI 2030'' Major Project (2018AAA0100900) and National Natural Science Foundation of China (62177033). The work is also sponsored by Huawei Innovation Research Program. We thank
MindSpore~\cite{mindspore}, a new deep learning computing framework, for the partial support of this work.
\end{acks}

\balance
\bibliographystyle{ACM-Reference-Format}
\bibliography{MIR}

\appendix
\section{Proof of Proposition \ref{pro:slattention}}\label{appe}
SLAttention is composed of the affinity matrix, the personalized interest decay, and the final attention, so we prove the permutation-equivariance of these parts in turn.  To start with, let $IA(\cdot)$ be the item-level interaction function defined in Eq.\eqref{eq:AI}, and with any permutation $\pi\in \Pi_n$ applying to input $\mathbf{S}=\{\mathbf{s}_1,\mathbf{s}_2,...,\mathbf{s}_n\}$, we get
\begin{equation}
    IA(\pi \mathbf{S})= IA([\mathbf{s}_{\pi(i)}]_{\pi(i)}) = [\mathbf{s}_{\pi(i)} W_{IA}L]_{\pi(i)}=\pi IA(\mathbf{S})\,,
\end{equation}
where $\pi(i)$ is the $i$-th element in permutation $\pi$ and $[\mathbf{s}_{\pi(i)}]_{\pi(i)}$ is a permutation of $\mathbf{S}$ ordered by $\pi(i)$, $i=1,\ldots,n$. 
Let $FA(\cdot)$ be the feature-level interaction function in Eq. \eqref{eq:FA}.
Since this interaction only involves feature interaction between items and does not change the original order of items, we can obtain
\begin{equation}
    FA(\pi \mathbf{S})+IA(\pi \mathbf{S})=\pi(FA(\mathbf{S})+IA(\mathbf{S}))\,.
\end{equation}

As for personalized interest decay, the decay matrix $D\in \mathbb{R}^{n\times m}$ in Eq. \eqref{eq:decaymat} consists of $n$ identical vectors $\mathbf{d}=\{d_1,d_2,...d_m\}$ where $d_i$ denotes the decay weight for the $i$-th items in the history list. Thus, the decay is distinct for items in the history list and identical for items in the candidate set. Let $PD(\cdot)$ and $\mathbf{C}_A=\{\mathbf{c}^a_1,\mathbf{c}^a_2,...,\mathbf{c}^a_n\}$ be the process of personalized interest decay and its input, thus 
\begin{equation}
\begin{split}
    PD(\pi \mathbf{C}_A)&=PD([\mathbf{c}^a_{\pi(i)}]_{\pi(i)})=[\mathbf{c}^a_{\pi(i)}]_{\pi(i)}\odot(\mathbf{E}+\mathbf{D})\\
    &=[\mathbf{c}^a_{\pi(i)}\odot(\mathbf{e}+\mathbf{d})]_{\pi(i)}=\pi PD(\mathbf{C}_A)=\pi\mathbf{C}\,,
\end{split}
\end{equation}
where $\mathbf{E}\in\mathbb{R}^{n\times m}$ and $\mathbf{e}\in\mathbb{R}^{1\times m}$ denote the unit matrix and its corresponding uint vector. The matrix $\mathbf{C}$ is the result of personalized interest decay obtained in Eq \eqref{eq:PID}.

For the final attention part in Eq. \eqref{eq:att1}, the attention coefficient $\mathbf{A}_S$ for the candidate set can be written as follows,  
\begin{equation}
    \begin{split}
        \mathbf{A}_S&=\text{softmax}\big(\mathbf{S}\mathbf{W}_S+\mathbf{C}(\mathbf{L}\mathbf{W}_L)\big)\\
        &=\text{softmax}\big([\mathbf{s}_i\mathbf{W}_S+\mathbf{c}_i(\mathbf{L}\mathbf{W}_L)]_i\big)\\
        &=\big[\text{softmax}(\mathbf{s}_i\mathbf{W}_S+\mathbf{c}_i(\mathbf{L}\mathbf{W}_L))\big]_i\,.\\
    \end{split}
\end{equation}
Thus, with $\pi$ applied to $\mathbf{S}$ and $\mathbf{C}$, we can get $\pi\mathbf{A}_S$. Then, the permutation-equivariance of the output $\hat{\mathbf{S}}$ in Eq. \eqref{eq:att2} can be proved by
\begin{equation}
\begin{split}
    (\pi\mathbf{A}_S)(\pi\mathbf{S})&=\Big[\sum\nolimits_j \mathbf{A}_S(\pi(i),\, \pi(j))\mathbf{S}_{\pi(j)}\Big]_{\pi(i)}\\
    &=\Big[\sum\nolimits_j \mathbf{A}_S(\pi(i),\, j)\mathbf{S}_{j}\Big]_{\pi(i)}
    =\pi(\mathbf{A}_S\mathbf{S})=\pi\hat{\mathbf{S}}\,.
\end{split}
\label{eq:fin-att}
\end{equation}
The attention coefficient $\mathbf{A}_L$ can be written as  
\begin{equation}
    \begin{split}
        \mathbf{A}_L&=\text{softmax}(\mathbf{S}\mathbf{W}_S\mathbf{C})
        =\text{softmax}\Big(\Big[\sum\nolimits_j\mathbf{s}_i\mathbf{w}^s_j\cdot\mathbf{c}_j\Big]_i\Big)\\
        &=\Big[\text{softmax}(\sum\nolimits_j\mathbf{s}_i\mathbf{w}^s_j\cdot\mathbf{c}_j)\Big]_i\,.\\
    \end{split}
\end{equation}
Similar to Eq. \eqref{eq:fin-att}, we have $\pi\mathbf{A}_L$ with $\pi$ applied to $\mathbf{S}$ and $\mathbf{C}$. Since $\mathbf{L}$ has nothing to do with candidate set, it does not harm the permutation-equivariance to multiply $\mathbf{A}_L$ and $\mathbf{L}$ in Eq. \eqref{eq:att2}. 

Combining all we proved above, we can conclude that SLAttention structure is permutation-equivariant to the candidate set.

\section{Details of Metrics}\label{metrics}
According to \cite{dendcg}, for linearly decomposable ranking metrics like DCG, an unbiased estimation can be obtained via IPS weighting:
 \begin{equation}
     \delta(\pi|u) = \sum_{v\in \pi}\frac{\lambda(pos(v|\pi))y_v}{Prop(o_u(v)=1|u, v, \pi^h)}
 \end{equation}
where $\pi$ and $\pi^h$ denote the current top-$K$ ranking and the ranking in the log data, and $\lambda(\cdot)$ can be any weighting function that depends on the position $pos(v|\pi)$ of item $v$ in ranking $\pi$. The label $y_v$ denotes whether the user $u$ clicked item $v$ in the log data. $Prop(o_u(v)=1|u, v,\pi^h)$ describes the propensity of user $u$ observing item $v$ in historical ranking $\pi^h$. Here, following \cite{u-rank}, we adopt category-wise propensity estimation as a coarse approximation of the groundtruth propensity $Prop(o_u(v)=1|u, v, \pi^h)$. We count the number of clicks on each category at each position and then normalize them via the number of clicks on the first position to obtain the category-wise propensity $Prop(o_u(v)=1|\text{cate}_v, \text{pos}_v)$, where $\text{cate}_v$ and $\text{pos}_v$ are the category and position of item $v$. With $\lambda(pos(v|\pi))=\frac{1}{\log_2(pos(v|\pi)+1)}$, we have the unbiased \textit{DCG}, which is normalized by the ideal unbiased \textit{DCG} to obtain \textit{deNDCG@K}. 

we also use IPS weighting to debias the utility metrics. On public datasets, $Utility@K$ is computed by $\sum_{v=1}^K\frac{Prop(o_u(v)=1|\text{cate}_v, \text{pos}_v)}{Prop(o_u(v)=1|\text{cate}_v, \text{pos}_v^h)}y_v$, following \cite{u-rank}, where $\text{pos}_v^h$ and $y_v$ denote the position of item $v$ and whether it was clicked in log data. On App Store dataset, $Utility@K$ is calculated by $\sum_{v=1}^K\frac{Prop(o_u(v)=1|\text{cate}_v, \text{pos}_v)}{Prop(o_u(v)=1|\text{cate}_v, \text{pos}_v^h)}y_v\gamma_v$, where $\gamma_v$ is the given bid price for item $v$. 
\end{document}